\documentclass[conference]{IEEEtran}
\usepackage{amsmath,amssymb,amsfonts,amsthm}
\usepackage{algorithmic}
\usepackage{textcomp}
\usepackage{color}
\usepackage{xcolor}
\usepackage[ruled,linesnumbered]{algorithm2e}
\usepackage{multirow}
\usepackage{makecell}
\usepackage{booktabs}
\usepackage{outlines}
\usepackage{bbold}
\usepackage{graphicx}
\usepackage{subfigure}
\usepackage{newtxtext}
\usepackage{fancyhdr}
\usepackage{enumitem}
\usepackage{soul}
\usepackage{color}
\usepackage{comment}
\usepackage{diagbox}
\usepackage{epstopdf}
\usepackage{wrapfig}
\usepackage{pifont}
\usepackage[nocompress]{cite}
\usepackage[hidelinks,colorlinks=true,allcolors=blue]{hyperref}

\usepackage{algorithmic}
\usepackage{textcomp}
\usepackage{orcidlink}

\definecolor{darkgreen}{rgb}{0.0, 0.5, 0.0}

\newcommand{\eg}{{\it e.g.}}

\newcommand{\ie}{{\it i.e.}}

\newcommand{\sys}{\textsc{FedCod}\xspace}

\IEEEoverridecommandlockouts
\usepackage{cite}
\usepackage{amsmath,amssymb,amsfonts}
\usepackage{algorithmic}
\usepackage{graphicx}
\usepackage{textcomp}
\usepackage{xcolor}
\def\BibTeX{{\rm B\kern-.05em{\sc i\kern-.025em b}\kern-.08em
    T\kern-.1667em\lower.7ex\hbox{E}\kern-.125emX}}
\begin{document}

\title{\sys: An Efficient Coded Communication Protocol for Cross-Silo Federated Learning}


\author{
Peishen Yan$^{\dagger}$,
Jun Li$^{\ddagger}$,
Hao Wang$^{\S}$,
Yang Hua$^{\P}$,
Tao Song$^{\dagger}$,
Lu Peng$^{\|}$,
Haibing Guan$^{\dagger *}$\\[0.5em]
$^{\dagger}$Shanghai Jiao Tong University, 
$^{\ddagger}$City University of New York, 
$^{\S}$Stevens Institute of Technology,\\
$^{\P}$Queen's University Belfast, 
$^{\|}$Tulane University\\
\{peishenyan, songt333, hbguan\}@sjtu.edu.cn,
jun.li@qc.cuny.edu,\\
hwang9@stevens.edu,
y.hua@qub.ac.uk,
lpeng3@tulane.edu
\thanks{
$^{*}$\textit{Corresponding author}
}
}

\maketitle
\thispagestyle{fancy}
\lfoot{979-8-3195-1703-6/26/\$31.00~\copyright~2026 IEEE}
\cfoot{}
\renewcommand{\headrulewidth}{0mm}

\begin{abstract}
Federated Learning (FL) is an innovative distributed machine learning paradigm that enables multiple parties to collaboratively train a model while preserving data privacy. 
Communication efficiency concerns arise in cross-silo FL, particularly due to the network heterogeneity and fluctuations associated with geo-distributed data silos. Most of the existing solutions to these problems focus on algorithmic improvements that alter the FL algorithm but sacrifice the training performance.
How to address these problems from a network perspective that is decoupled from the FL algorithm remains an open challenge.
This paper proposes \sys, the first application-layer coded communication protocol designed for cross-silo FL. \sys transparently utilizes a coding mechanism to enhance the efficient use of idle bandwidth through client-to-client communication, and dynamically adjusts coding redundancy to mitigate network bottlenecks and fluctuations, thereby improving the communication efficiency and accelerating the FL training process. 
In our real-world experiments, \sys demonstrates a significant reduction in average communication time by up to 62\% compared to the baseline, while maintaining FL training performance and optimizing inter-client communication traffic.

\end{abstract}


\section{Introduction}\label{sec:intro}
Unlike cross-device federated learning (FL) that distributes model training to loosely connected devices (\eg, Internet-of-Things devices and smart phones)~\cite{konevcny2015federated,li2020federatedmlsys}, cross-silo FL performs distributed training across geo-distributed data centers (known as ``silos'') powered with sufficient computing resources, connected by wide-area networks (WANs), and provisioned with large amounts of data~\cite{kairouz2021advances, marfoq2020throughput}. 
Cross-silo FL has been widely applied to enable collaborative training between organizations to build artificial intelligence (AI) models without sharing their private data, such as healthcare companies~\cite{oldenhof2023industry, silva2019federated,courtiol2019deep} and financial institutions~\cite{liu2021fate}. 
The release of the EU AI Act~\cite{madiega2021artificial}, the popularity of large language models (LLMs), and scarcity of high-quality data for LLM training will jointly boost the development and deployment of cross-silo FL~\cite{woisetschlager2024federated,ye2024openfedllm}.

However, cross-silo FL has been suffering from real-world network conditions, leading to inflated training time and inefficient utilization of client data: \textbf{1) Increasing bandwidth demand}:  
With explosively increasing neural network model sizes~\cite{fan2023fate,zhang2024towards}, particularly LLMs~\cite{kuang2024federatedscope}, the limited WAN bandwidth has been throttling cross-silo FL's training efficiency~\cite{hsieh2017gaia}. 
\textbf{2) Network heterogeneity}: The geo-distributed nature of cross-silo FL leads to heterogeneous connections between silos, creating bottlenecks that extensively slow down the exchange of model weights~\cite{hsieh2017gaia,sattler2019robust,ye2023heterogeneous}. 
\textbf{3) Bandwidth fluctuations}:  Variability in WAN traffic over public links leads to fluctuating bandwidth over time~\cite{saeed2020annulus,wang2021examination}, exacerbating training inefficiencies by increasing transmission delays. 

Most existing methods address these problems through algorithmic innovation, either by performing model compression~\cite{rothchild2020fetchsgd,dorfman2023docofl} or by refactoring training and aggregation strategies~\cite{li2024accelerating,luo2021cost,ouyang2022clusterfl}.  
However, such algorithmic solutions typically trade off between model quality and training efficiency, which may result in sub-optimal performance under highly volatile cross-silo FL networks~\cite{karimireddy2020scaffold,liu2022hierarchical,sun2024convergence,sun2023mode}. 
This dilemma inspires researchers to further explore solutions from a network perspective~\cite{abad2020hierarchical,wang2021resource,xu2022hierfedml}. 
For example, HierFL~\cite{wang2021resource} organizes clients into clusters and performs hierarchical model aggregation based on network topology and geographic proximity to reduce bandwidth consumption and improve communication efficiency. 
However, hierarchical aggregation still requires fast and stable connections between clients within the same cluster. 
Consequently, clustering-based solutions struggle to adapt cross-silo FL networks to the heterogeneous and fluctuating conditions of WANs due to the sparsity of silos and their geo-distributed nature. 
Therefore, \textit{optimizing communication efficiency in cross-silo FL without sacrificing model quality remains an open challenge}.

This paper proposes \sys, the first application-layer \textit{coded} communication protocol tailored to geo-distributed cross-silo FL, which integrates coding into both download and upload phases to optimize communication efficiency and accelerate FL training.
Specifically, \sys adapts to heterogeneous and fluctuating networks transparently by employing dynamic coding redundancy and coded aggregation strategies.

The coded communication protocol, at the core of \sys, has been seamlessly integrated into the whole life cycle of cross-silo FL to address the challenges of WAN communications.  
To mitigate the impact of limited WAN bandwidth between data silos, \sys encodes the model weights into redundant data blocks, structuring multiple transmission paths between sources and sinks through client-to-client communication, utilizing the idle bandwidth in cross-silo FL networks.
The coding redundancy enables cross-silo FL to tolerate heterogeneous and bottleneck links by adaptively selecting faster links to retrieve encoded data blocks. This mechanism ensures that model weights can be efficiently recovered without being blocked by slow or faulty links. 
Besides, \sys allows clients to forward encoded data blocks from their neighbors to the server and pre-aggregate with their own blocks. The client-side pre-aggregation further reduces the upward bandwidth consumption to the server. 
Meanwhile, \sys's adaptive redundancy algorithm establishes a minimum level of coding redundancy to ensure resilience against network fluctuations. \sys gradually reduces the redundancy when performance is stable. When performance fluctuations are detected, \sys promptly restores high redundancy to maintain efficient communication, effectively avoiding network bottlenecks and fluctuations in cross-silo FL networks. 

Our main contributions are as follows:

\begin{itemize}[noitemsep,topsep=0pt,parsep=0pt,partopsep=0pt]
\item We are the first to introduce application-layer \textit{coded} communication protocol tailored to geo-distributed cross-silo FL, which addresses network fluctuations and heterogeneity to enhance collaborative training efficiency from a network perspective.

\item We propose \sys, an application-layer coded communication protocol that leverages adaptive coding redundancy to enhance cross-silo FL communication efficiency while reducing traffic. 

\item We empirically evaluate \sys in diverse real-world cross-silo environments, showing that it reduces communication overhead by up to 62\% compared to the baseline protocol while preserving FL training performance and optimizing inter-client communication traffic.
\end{itemize}

\section{Background \& Motivation}
\subsection{Cross-Silo FL}
\label{sec:notation}

According to FL participant types and distribution, FL can be categorized into \textit{cross-device} and \textit{cross-silo} FL. 
The cross-device FL scenario involves a large number of clients typically with limited processing power and unreliable communication links~\cite{kairouz2021advances}.
Unlike cross-device FL, cross-silo FL participants are usually organizations (\eg, medical institutions) with sufficient computing resources, connected by WANs,
typically housed in geo-distributed data centers~\cite{kairouz2021advances}.  

A cross-silo FL system involves a server and $n$ clients.\footnote{The ``server'' and ``clients'' are logical roles in cross-silo FL. The ``server'' indicates the organization who orchestrates the learning and aggregates weights. The ``clients'' are organizations performing training on local data.} 
Each client $i$ possesses a local dataset $D_i$. 
In each communication round $t$, the activities of the server and the clients can be divided into the following phases: 
1) \textit{Download phase}: The server broadcasts the global model $W_t$ to all clients. The download time for client $i$ is denoted as $T_\text{download}(i)$. 
2) \textit{Training phase}: Each client $i$ receives the global model $W_t$ as the initial model $W^i_{t+1}$, and performs several iterations of local training on its local dataset. Client $i$'s training time is defined as $T_\text{train}(i)$. 
3) \textit{Upload phase}: After local training, each client $i$ uploads its model weights $W_{t+1}^i$ to the server, which has a single sink and multiple sources. The upload time is $T_\text{upload}(i)$.
4) \textit{Aggregation phase}: The server aggregates the received local models using a predefined aggregation algorithm. For example, FedAvg~\cite{mcmahan2017communication} and FedProx~\cite{li2020federatedmlsys} use a weighted average to update the global model: $W_{t+1} = \sum_{i} \frac{|D_i|}{\sum_{i}|D_i|} W_{t+1}^i$. Many other algorithms also aggregate linearly~\cite{li2023revisiting, deng2022improving}.

The total time $T(i)$ from the start of the download phase to the receipt of the local model from client $i$ at the server is given by $T(i) = T_\text{download}(i) + T_\text{train}(i) + T_\text{upload}(i)$. 
Therefore, the duration of one communication round is $T = \max_i T(i)$, and client $i$'s waiting time is defined as $T_\text{wait}(i)=T-T(i)$.

\begin{figure}[t]
    \centering
    \subfigure[Global profiling results]{\includegraphics[width=0.24\textwidth]{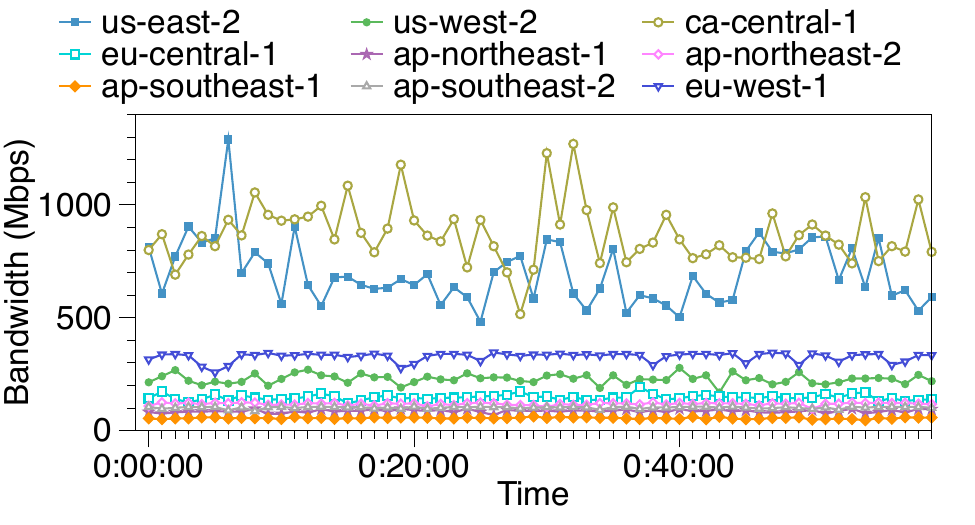}\label{fig:fluct-aws}}
    \subfigure[North America profiling results]{\includegraphics[width=0.24\textwidth]{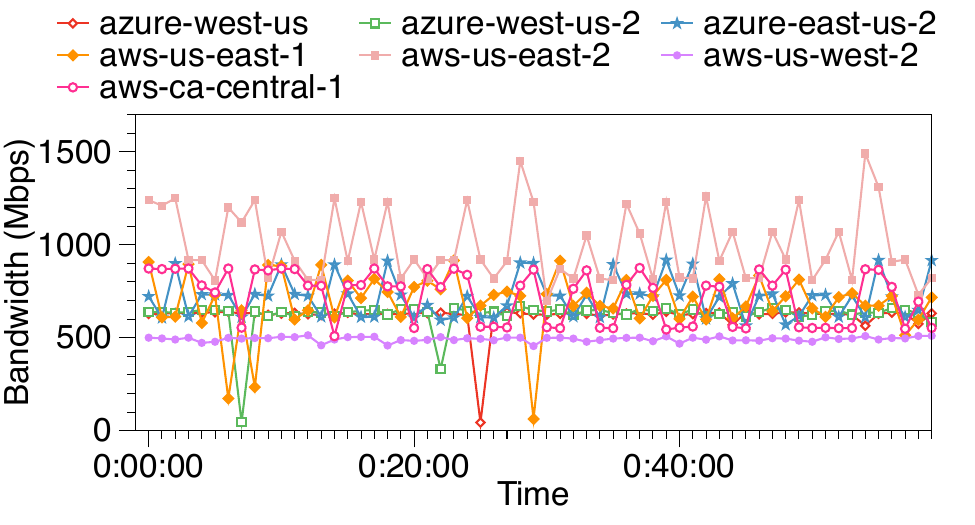}\label{fig:fluct-all}} 
    \vspace{-0.15in}
    \caption{Communication bandwidth profiling results.}
    \vspace{-0.2in} 
    \label{fig:profile}
\end{figure}

\subsection{Measuring Real-world Cross-Silo Networks}

In real-world cross-silo FL scenarios, clients experience prolonged communication due to limited and fluctuating WAN bandwidth, even though equipped with high-speed network interfaces and broadband Internet access.
To understand and quantify the challenges in cross-silo FL, we measure cross-silo FL networks and evaluate the issues highlighted above: \textit{network heterogeneity} and \textit{bandwidth fluctuations}.

We used \texttt{iPerf} to profile the communication bandwidth between the server and clients over an hour for both regional (North America) and global network topologies of cross-silo FL. 
More experimental details refer to Section~\ref{sec:exp-static}.
The results in Figures~\ref{fig:fluct-aws} and \ref{fig:fluct-all} illustrate the challenges of \textit{network heterogeneity} and \textit{bandwidth fluctuations}.

\emph{Spatially}, client-server bandwidth varies widely from tens of megabits per second to over 1 Gbps, reflecting severe \emph{network heterogeneity} largely driven by geographic distance (\eg, intercontinental links). Such slow links create stragglers that bottleneck synchronous FL. \emph{Temporally}, per-link bandwidth is highly volatile and may drop to near zero, so large model transfers can take minutes and experience multiple bandwidth peaks and valleys, making scheduling difficult.

These observations indicate that cross-silo FL requires communication schemes robust to both heterogeneous and time-varying WAN bandwidth.

\subsection{Motivating Coded Communication} 
\label{sec:motivation}
\begin{figure}[h]
        \centering
        \vspace{-0.15in}
        \includegraphics[width=0.38\textwidth]{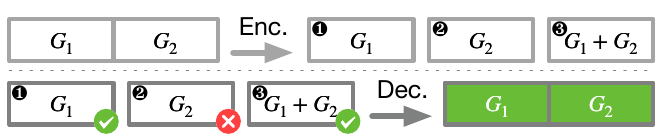}
        \vspace{-0.05in}
        \caption{A simple example for coding.}
        \label{fig:coding-example}
        \vspace{-0.1in}
\end{figure}

Coding provides space-efficient redundancy, enabling systems to recover from failures and bottlenecks by reconstructing lost data from redundant pieces~\cite{wu2006echelon,feng2008large,huang2012erasure}. 
As illustrated in Figure~\ref{fig:coding-example}, we can recover the data $G=[G_1,G_2]$ from the encoded data blocks 1 and 3, even though block 2 is lost.

In cross-silo FL, bandwidth is often underutilized: while some client-server links are bottlenecked, other client-to-client paths remain idle. This creates an opportunity to exploit multi-path transfers, where model blocks can be relayed across clients and recovered from a subset of received blocks via coding, making communication robust to heterogeneous and time-varying WAN bandwidth.

\begin{figure}[t]
    \centering
    \subfigure[The communication overhead for baseline and our two adapted network coding communication protocols (one in download phase and the other in upload phase).]{\includegraphics[width=0.21\textwidth, trim=0 0 0 5]{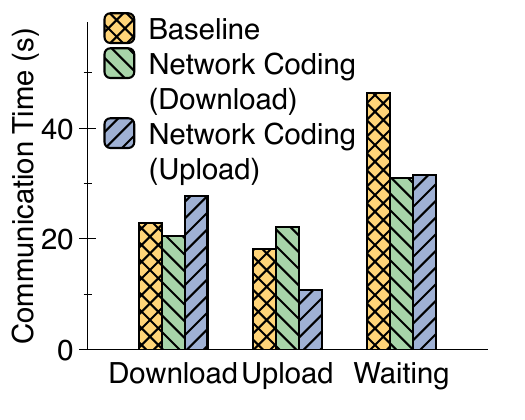}\label{fig:example}}    
    \subfigure[Bandwidth and upload workload distribution between different clients and server. (Left) Bandwidth distribution. (Right) Proportion of blocks received from each client during upload.]{\includegraphics[width=0.27\textwidth, trim=0 0 0 5]{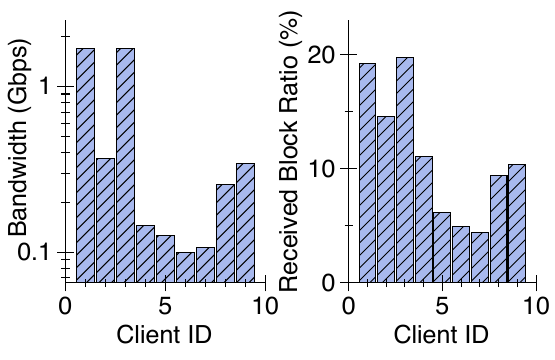}\label{fig:example-explain}}
    \caption{Motivating Example.}
\end{figure}

To motivate our design, we run a real-world experiment using random linear network coding (RLNC)~\cite{ho2006random} in both the global download and local upload phases. As shown in Figure~\ref{fig:example}, RLNC reduces straggler waiting in the download phase by completing clients' downloads more synchronously, and cuts both upload time and waiting time by 30\% in the upload phase.
Intuitively, coding redundancy lets faster links contribute more blocks without being gated by slower links (Figure~\ref{fig:example-explain}). Motivated by these results, we design coding strategies tailored to cross-silo FL dataflow to adapt to network heterogeneity and fluctuations.

\section{Overview}

\subsection{Objectives \& Challenges}

\sys is designed to achieve the following objectives:

\textbf{O1: Accelerating FL by optimizing the communication efficiency.} \sys aims to accelerate FL by optimizing communication efficiency with coded communication protocols that fully utilize cross-silo FL networks' available bandwidth.

\textbf{O2: Adapting to fluctuating network conditions.}
As network conditions across geo-distributed data silos exhibit spatial network heterogeneity and temporal bandwidth fluctuations, \sys should be able to adapt to varying network conditions by carefully applying coding techniques. 

\textbf{O3: Compatibility with existing FL algorithms.}
As an application-layer coded communication protocol, the design should be compatible with existing FL algorithms.

To achieve these objectives, we should address the corresponding challenges:

First, \emph{designing effective coding strategies for cross-silo FL is non-trivial}. Coding must balance redundancy and latency mitigation: insufficient redundancy reduces robustness under heterogeneity and fluctuations, while excessive redundancy wastes bandwidth and increases overhead. Moreover, centralized FL can bottleneck at the server due to heavy ingress and egress traffic. Thus, \sys should optimize coding to improve both fault tolerance and communication efficiency while reducing server traffic.

Second, \textit{the variety of network conditions} make it hard to identify optimal transmission paths and balance load across clients. 
Leveraging idle bandwidth via client-to-client transfers introduces additional paths and can improve efficiency. However, static optimization policies quickly become suboptimal under bandwidth fluctuations, and real-time global profiling is often infeasible, complicating online adaptation.

Lastly, \emph{existing communication-efficient FL schemes are often tightly coupled with specific FL algorithms} (\eg, aggregation strategies or communication frequency), which hinders integration and limits generality. We therefore seek a communication protocol that can be plugged into different cross-silo FL algorithms with minimal assumptions.

\begin{figure}[t]
    \centering
    \subfigure[Download]{\includegraphics[width=0.38\textwidth,trim=0 0 0 0]{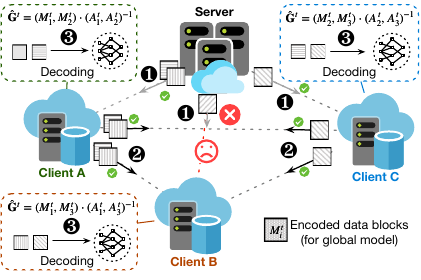}\label{fig:arch-download}}\\
    \subfigure[Upload]{\includegraphics[width=0.38\textwidth,trim=0 0 0 0]{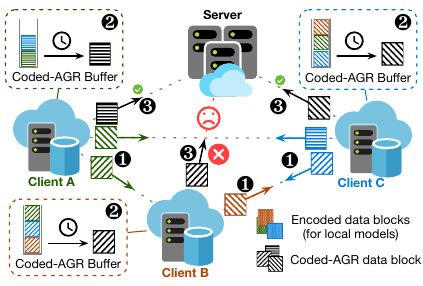}\label{fig:arch-upload}}   
    \caption{\sys's workflow. (Illustrated with the number of model partitions $k=2$ as an example.)}
    \label{fig:arch}
\end{figure}

\subsection{\sys's Workflow}

\sys improves communication efficiency in cross-silo FL by deploying coded transfer to the \textit{Download} and \textit{Upload} phases. Both the server and clients perform lightweight coding in addition to standard transmission, while \textit{Training} and \textit{Aggregation} remain unchanged.

In the \textit{Download} phase (Figure~\ref{fig:arch-download}), the server encodes model partitions with random coefficients: \textbf{Step} \ding{182}: The server sends distinct encoded data blocks to clients; \textbf{Step} \ding{183}: Each client forwards received data blocks to its neighboring clients; \textbf{Step} \ding{184}: A client decodes the global model once enough encoded data blocks are collected.

The \textit{Training} phase follows the standard procedure in which each client performs local training and updates its model. 

In the \textit{Upload} phase (Figure~\ref{fig:arch-upload}), clients encode local updates using a shared coefficient sequence: \textbf{Step} \ding{182}: The clients send the encoded data blocks to designated neighbors according to the predefined mapping; \textbf{Step} \ding{183}: Each client maintains a Coded-AGR buffer and aggregates the encoded data blocks with the same coefficient vector into an AGR data block; \textbf{Step} \ding{184}: The clients send the AGR data blocks to the server, which decodes the aggregated global model.

After \textit{Aggregation}, an adaptive redundancy algorithm updates coding parameters to improve efficiency and traffic.

\section{\sys's Design}

We introduce \sys, an application-layer coded communication protocol for centralized cross-silo FL to improve communication efficiency. 
\sys consists of \textbf{coding strategies} and an \textbf{adaptive redundancy algorithm}.
By applying different coding strategies for distinct data flow characteristics of the download and upload phases, \sys optimizes bandwidth utilization and accelerates the transmission of model weights through client-to-client communication. 
Increased coding redundancy improves communication efficiency and system tolerance to network fluctuations, but more redundancy than necessary results in wasted communication traffic.
The adaptive redundancy algorithm dynamically adjusts the level of coding redundancy to changing network conditions, balancing increased tolerance with increased communication traffic.

\subsection{Coding for the Download Phase}\label{sec:nc-download}

In the download phase, the server distributes the global model to all clients, which is a single-source content distribution task. The basic server-client communication protocol results in repetitive transmission of identical data from the server to multiple clients, creating a bottleneck at the server due to the high volume of egress traffic.

\noindent\textbf{Server-side Encoding}. To reduce the transmission of duplicate data, the server splits the global model into multiple partitions of equal size and encodes these partitions into encoded data blocks before sending them to the clients. 
Specifically, the global model $\mathbf{G}^t$ is divided into $k$ equal-sized partitions at each communication round, denoted as $\mathbf{G}^t=(G_1^t, G_2^t,\dots, G_k^t)$, where $t$ indicates the $t$-th communication round. New encoded blocks $\{M_i^t\}$ are continuously generated by different linear combinations of the model partitions with the random coefficient vector $A_i^t=(\alpha_{i,1}^t, \alpha_{i,2}^t,\dots, \alpha_{i,k}^t)^T$, \ie,
\begin{equation}
    M_i^t = \mathbf{G}^t\cdot A_i^t = \sum_{j=1}^k \alpha_{i,j}^t \cdot G_j^t.
\end{equation}
The server generates the coefficient vector $A_i^t$ and appends it to the block header. It sends encoded data blocks to clients until each can decode the global model.
Since coding cost grows with the number of blocks~\cite{li2011random}, it is necessary to bound the number of partitions $k$.

\noindent\textbf{Client-side Decoding}. According to coding theory, if a client has received $k$ different data blocks $\{M_{i_j}^t\}_{j=1,\dots,k}$ with the corresponding coefficient vectors $\mathbf{A}=(A_{i_1}^t,A_{i_2}^t,\dots,A_{i_k}^t)$, these vectors are linearly independent with a high probability close to 1. The client can recover the original $k$ global model partitions with Gaussian elimination as the following equation:
\begin{equation}
\begin{aligned}
    \hat{G}^t_i = \sum_{j=1}^k \mathbf{A}^{-1}[i][j]\cdot M^t_{i_j}, \text{ for }i=1,\dots,k,\\
    \text{\ie, } \hat{\mathbf{G}}^t=(M_{i_1}^t,\dots,M_{i_k}^t)\cdot \mathbf{A}^{-1}.
\end{aligned}
\end{equation}

\noindent\textbf{Client-assisted Forwarding}.\label{sec:forwarding} 
The randomness of the coding coefficients ensures that any $l$ ($l\le k$) encoded blocks sent by the server are linearly independent. The clients can leverage the available client-to-client links to forward blocks received from the server to neighboring clients. Therefore, under this strategy, each client only needs to receive any $k$ encoded blocks to achieve the global model, regardless of which communication link they come from. The server significantly reduces egress traffic, and the client benefits from faster communication links while reducing its dependence on a single channel. 

Distinct from network coding, we do not perform client-side re-encoding and we do not forward data blocks received from other clients, due to the fact that they would result in linear dependency of the forwarded data blocks~\cite{wang2006practical}. In contrast, our forwarding strategy ensures that there is no duplicate data transmission by linearly independent coding, and eliminates the overhead of client-side re-encoding and memory copying of data blocks, making communication more efficient.

\subsection{Coding for the Upload Phase}\label{sec:nc-upload}
In the upload phase, we adopt a coding and forwarding strategy similar to that of the download phase to improve communication efficiency. Each client encodes its local model into different data blocks and sends them to the server and to neighboring clients. The clients then forward the data blocks received from their neighbors to the server. The server decodes and aggregates the local models from the received data blocks.

To efficiently manage the transmission process, each client prioritizes uploading its own data blocks. Specifically, each client maintains two upload queues: \textit{own-queue} and \textit{other-queue}. The \textit{own-queue} is dedicated to the uploading of $k$ encoded data blocks of its local model. The \textit{other-queue} is used for forwarding the encoded data blocks from other clients. If the \textit{own-queue} is not empty, the client prioritizes sending blocks in this queue to the server; otherwise, the client sends data blocks from the \textit{other-queue} to the server in a first-in, first-out (FIFO) order.

\subsection{Coded Aggregation}\label{sec:preagr}
Unlike the download phase, the upload phase involves multiple sources sending data to a single sink, resulting in heavier and more complex data flow. 
The amount of information that the server needs to receive is $n$ times the amount sent during the download phase and grows linearly with the number of clients, which can make the server a bottleneck as $n$ increases.
At the same time, the coding redundancy increases exponentially, contributing further to the network load. Therefore, addressing the network issues through coding alone is challenging.

Considering that coding works linearly on model partitions and most existing aggregation algorithms are linear, we design the Coded Aggregation (Coded-AGR) based on the coding strategy for upload in Section~\ref{sec:nc-upload}. By aggregating encoded data blocks on the client side, Coded-AGR allows clients' local model partitions to hitchhike on other clients' local model partition uploads, which can reduce bandwidth occupancy during the transmission of data blocks. Benefiting from both coding and on-client aggregation, this approach improves network fluctuation tolerance and communication traffic savings.

During the upload phase, each client encodes its local model into different data blocks and transmits them to other clients. Then, on the client side, data blocks with the same coefficient vector are aggregated into an AGR data block that is forwarded to the server. Instead of reconstructing all individual local models, the server can reconstruct an aggregated global model from the AGR data blocks.

Specifically, all clients generate the same sequence of coefficient vectors required for coding as agreed upon in advance, \eg, based on the Cauchy matrix~\cite{cauchy1840exercices,plank2006optimizing}. 
For any two clients $i_1$ and $i_2$, the $j$-th generated coefficient vectors $(\alpha_{j,1}^{t,i_1}, \dots, \alpha_{j,k}^{t,i_1})$ are equal to $(\alpha_{j,1}^{t,i_2}, \dots, \alpha_{j,k}^{t,i_2})$. For simplicity, we denote the $j$-th coefficient vector as $\mathbf{\alpha}^t_j=(\alpha_{j,1}^{t}, \dots, \alpha_{j,k}^{t})$. Furthermore, they use the same mapping sequence $h(\cdot)$ to transmit the encoded data blocks to different clients, \ie, the $j$-th data block of the client $i$, $M_j^{t,i}$, is sent to the client $h(j)$.
For example, we define $h(j) = j\mod n$ as the default mapping.

When client $h(j)$ receives the data blocks $\{M^{t,i}_j\}$ from another client $i$, it aggregates them and its own $j$-th data block into a new data block $\tilde{M}^t_j=\sum_{\text{received }i} M^{t,i}_j$.
For client $h(j)$, it can choose to upload the data block $\tilde{M}^t_j$ once a predefined time window has elapsed (\textit{non-wait mode}), or it can upload the AGR data block after all the $j$-th blocks from all clients have been received and aggregated (\textit{wait mode}).

\textit{Wait mode} coded aggregation is theoretically more communication-efficient than \textit{non-wait mode}; we omit the proof due to space constraints. Unless stated otherwise, we use coded aggregation to refer to \textit{wait mode}.

\subsection{Adaptive Redundancy Algorithm}\label{sec:dynamic-r}
The improvement in communication efficiency through coding strategies relies on coding redundancy.
This redundancy enables clients with faster communication links to the server to assist neighboring clients in transmission, providing the system with tolerance to network fluctuations and link failures.
Inadequate redundancy can lead to less efficient communication, while excessive redundancy can lead to wasted communication traffic, which has a significant impact on client costs. To address this issue, we design the adaptive redundancy algorithm, which dynamically adjusts the coding redundancy to adapt to varying network conditions.

Overall, this algorithm responds to communication time variations caused by network fluctuations and sets a lower bound on redundancy to provide sufficient tolerance for common network fluctuations. The system starts with a high redundancy state, gradually reduces the redundancy if the communication time does not increase, and quickly restores high redundancy if performance fluctuations are detected.

Our adaptive redundancy algorithm consists of three parts: \textit{cold start}, \textit{redundancy reduction}, and \textit{rapid recovery}. 
\textit{1) Cold start}:
We define the number of encoded data blocks exceeding $k$ as the redundancy number $r$, and the coding redundancy of the system is defined as $r/k$. $r$ is initialized to a large number, indicating a high tolerance for network fluctuations.
\textit{2) Redundancy reduction}: The variables $t_\text{last}$ and $t_\text{cur}$ denote the communication duration of the last and the current communication round, respectively. $r$ is gradually reduced over time to reduce the redundancy of the communication volume. This reduction continues when the current communication duration $t_\text{cur}$ is not larger than the last communication duration $t_\text{last}$, indicating a decrease in redundancy demand. We also set a lower bound $r_\text{lb}$ for $r$ and adjust it according to the network conditions. To make the system less sensitive to small network fluctuations, we introduce a scaling factor $\lambda>1$ and the system compares $t_\text{cur}$ and $t_\text{last}*\lambda$ at the time of judgement. 
\textit{3) Rapid recovery}: If some links fail or bandwidth fluctuates, \ie, $t_\text{cur}$ exceeds $t_\text{last}*\lambda$, $r$ should be increased proportionally to ensure system resilience. The system should also increase the lower bound $r_\text{lb}$ because there is at least one communication path that is getting worse. The system will continue to adjust upwards $r$ in multiple rounds until the optimization of the system communication time is no longer apparent, \ie, $t_\text{cur} \ge t_\text{last} / \lambda$. In addition, if there are no further network fluctuations for a period of time, the system will decrease the lower bound of redundancy $r_\text{lb}$.

In essence, the algorithm dynamically adapts redundancy levels by starting with a high redundancy, gradually decreasing it to mitigate redundancy, and then increasing redundancy in response to network fluctuations. This iterative adjustment mechanism allows the system to optimize redundancy levels in accordance with evolving network conditions, thereby increasing overall efficiency and robustness.

It should be noted that adaptive redundancy is unnecessary in the download phase because transmissions are sequential per client and can be halted as soon as local decoding is possible, thus avoiding waste. In the upload phase, however, redundancy should be pre-arranged, making adaptive control essential.

\section{Evaluation}
This section answers the following questions: 1) What is the improvement in communication efficiency of the FL system using \sys compared to the basic server-client communication protocol? 2) What is the impact of \sys on the communication traffic? 3) Does \sys affect the learning convergence? 
\sys achieves up to 62\% reduction in communication time compared to the baseline and has the same convergence of the global model as the baseline.

\subsection{Experimental Setup}\label{sec:exp-setup}
We implement FL in PyTorch and use gRPC over TCP for communication.
For silo locations, we consider two topologies: Global and North America:

\noindent\textbf{Global topology}. We evaluate on Amazon Web Service (AWS) platform across 10 regions using one \texttt{p3.8xlarge} EC2 instance per region: \texttt{us-east-1} (N. Virginia), \texttt{us-east-2} (Ohio), \texttt{us-west-2} (Oregon), \texttt{ca-central-1} (Canada), \texttt{ap-northeast-1} (Tokyo), \texttt{ap-northeast-2} (Seoul), \texttt{ap-southeast-1} (Singapore), \texttt{ap-southeast-2} (Sydney), \texttt{eu-central-1} (Frankfurt), and \texttt{eu-west-1} (Ireland). Each instance has 32 vCPUs and up to 10~Gbps network bandwidth.

\begin{figure}[t]
    \centering
    \subfigure[In the global topology]{\includegraphics[width=0.23\textwidth,trim=0 0 0 0]{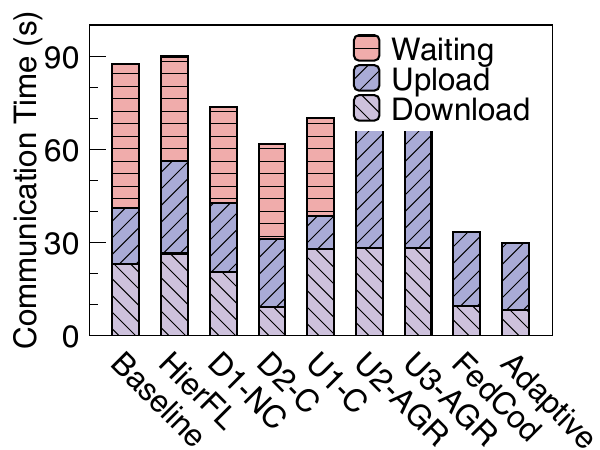}\label{fig:com-time-aws}}    
    \subfigure[In the North America topology]{\includegraphics[width=0.23\textwidth,trim=0 0 0 0]{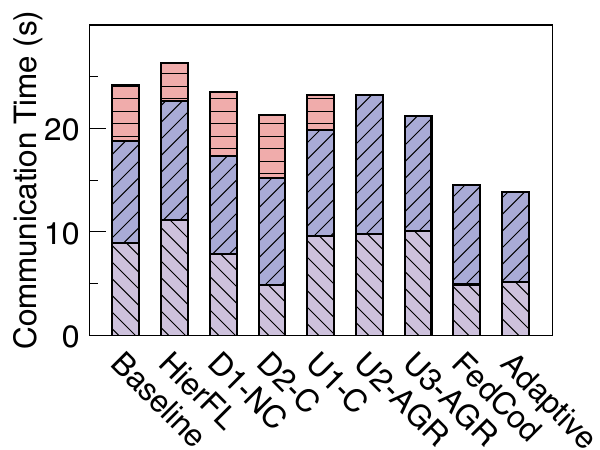}\label{fig:com-time-all}}
    \caption{The communication time of different protocols. Baseline: the basic server-client communication protocol; HierFL: the hierarchical FL communication protocol; D1-NC: network coding in the download phase; D2-C: our coding strategy only in the download phase; U1-C: our coding strategy only in the upload phase; U2-AGR: non-wait mode Coded-AGR in the upload phase; U3-AGR: wait mode Coded-AGR in the upload phase; FedCod: \sys with static redundancy; Adaptive: \sys with adaptive redundancy.}
    \vspace{-0.1in}
    \label{fig:com-time}
\end{figure}

\begin{figure*}[t]
    \centering
    \begin{minipage}[t]{0.68\textwidth}
    \vspace{0pt}
        \centering
        \includegraphics[width=\linewidth, trim=0 5 0 5]{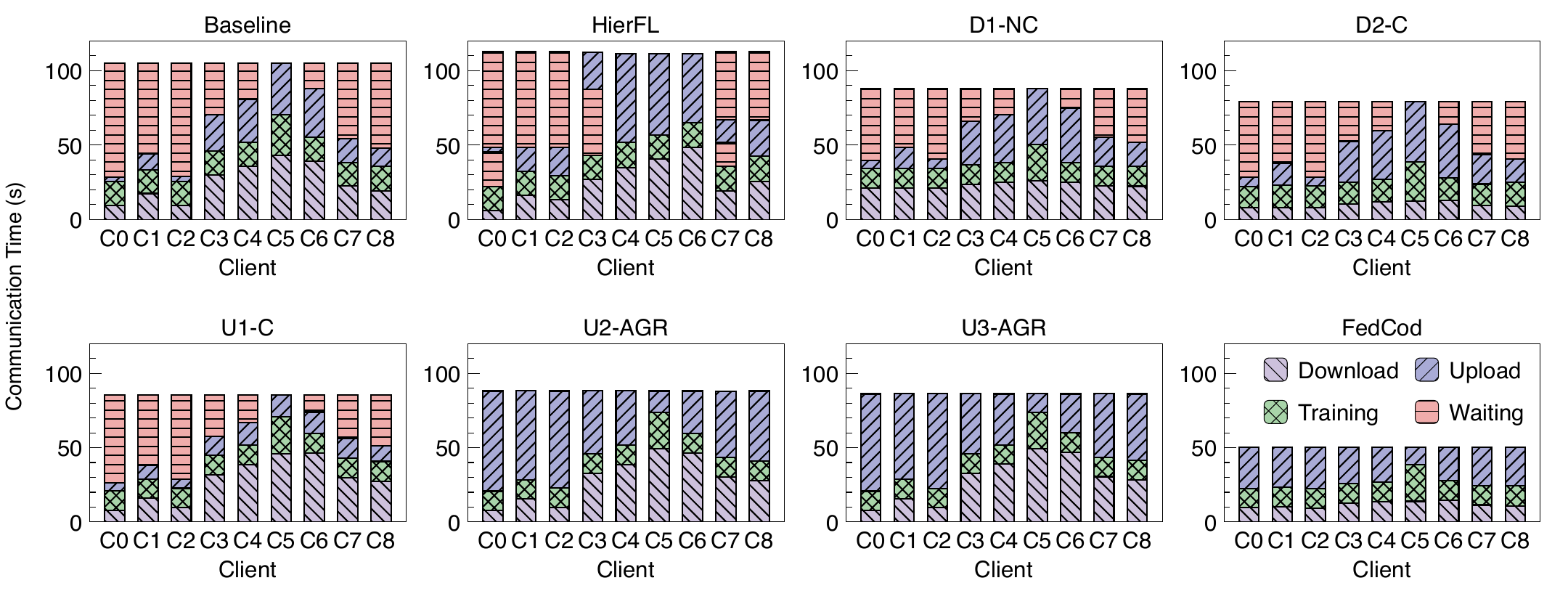}
        \vspace{-0.2in}
        \caption{The communication time for each client of different communication protocols.}
        \label{fig:split-time}
    \end{minipage}
    \hfill
    \begin{minipage}[t]{0.25\textwidth}
    \vspace{0pt}
        \centering
        \includegraphics[width=\linewidth, trim=0 10 0 5]{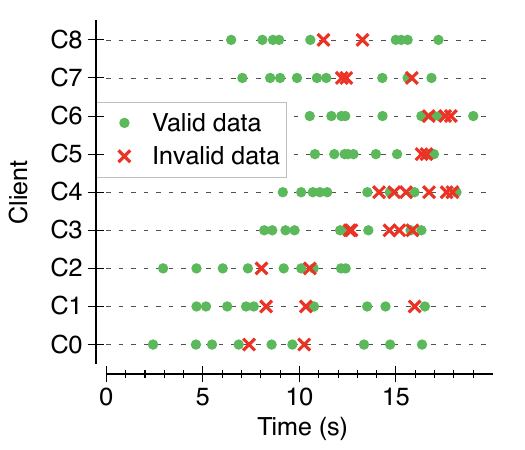}
    \vspace{-0.2in}
    \caption{Validity of encoded data blocks received on each client under network coding. The $x$-axis shows time since the server distributes the global model.}
    \label{fig:invalid-nc}
    \end{minipage}
    \vspace{-0.15in}
\end{figure*}

\noindent\textbf{North America topology}. We deploy on 8 North America regions: 4 AWS EC2 regions (\texttt{us-east-1}, \texttt{us-east-2}, \texttt{us-west-2}, \texttt{ca-central-1}) and 4 Azure regions (\texttt{central-us}, \texttt{west-us}, \texttt{west-us-2}, \texttt{east-us-2}). 
To reduce cost and focus on communication optimization, we sample client training time from an empirical distribution instead of performing GPU training.
On AWS, we use \texttt{m5.8xlarge} instances (32 vCPUs, comparable network interface). On Azure, we use \texttt{Standard\_D32a\_v4} instances (32 vCPUs, up to 16~Gbps), providing similar compute and network capacity.

In the global topology, we use AWS \texttt{us-east-1} as the server; the remaining instances are clients. In the North~America topology, we use Azure \texttt{central-us} as the server and the rest as clients.\footnote{We choose a geographically central region, following the setup in~\cite{marfoq2020throughput}.} 
We run 10 FL rounds per protocol and report the average phase time (Section~\ref{sec:notation}). Unless otherwise specified, the reported communication time is the end-to-end wall-clock time measured by the FL runtime, including data transfer as well as the additional protocol operations introduced by \sys, such as encoding, decoding, forwarding, and Coded-AGR. We also measure average ingress/egress traffic at the server and clients to quantify communication cost.
In addition, we conduct numerical studies on the number of model partitions and coding redundancy, and run conformance experiments on a local cluster to verify learning convergence.
We use ResNet152 (60M parameters; $\sim$440~MB packed weights) trained on federated datasets generated from the CIFAR-10 following~\cite{zeng2023fedlab}. Unless specified, we set the number of partitions $k=n$ and redundancy to 100\%.

\subsection{Experimental Results}

\subsubsection{Effectiveness of Coding Strategies}\label{sec:exp-static}
First, we show that Hierarchical FL (HierFL)~\cite{wang2021resource} is not suitable for the geo-distributed cross-silo FL scenario. We group data silos based on their geographical location and communication status and perform hierarchical aggregation within the group. In each group, the data silo with the fastest communication speed to the server is selected as the center for hierarchical aggregation. For example, in the global topology, the silos are divided into three groups, North America, Asia, and Europe. The centers of hierarchical aggregation are \texttt{us-east-2}, \texttt{ap-northeast-1} and \texttt{eu-central-1}. Figure~\ref{fig:com-time} shows that HierFL is even worse than the baseline.
With the more detailed composition of communication time per client in Figure~\ref{fig:split-time}, we can see that the communication time between clients within each group is non-negligible. As a result, forwarding most clients' data through the hierarchical aggregation center is less efficient than communicating directly with the server.

To demonstrate the effectiveness of each component of our \sys, we apply our designed coding strategies separately to the download and upload phases, and finally in combination, within the global topology. The adaptive redundancy algorithm of \sys is discussed in Section~\ref{sec:traffic-dynamic}; for now, we use static redundancy (with default redundancy 100\%).



\begin{table}[t]
\centering
\caption{The average server traffic (Unit: MBytes)}
\vspace{-0.05in}
\setlength{\tabcolsep}{2mm}
\begin{tabular}{ccc|ccc}
\hline
\multicolumn{3}{c|}{\textbf{Global}} & \multicolumn{3}{c}{\textbf{North America}} \\
\cline{1-3}\cline{4-6}
\textbf{Protocol} & \textbf{Ingress} & \textbf{Egress} &
\textbf{Protocol} & \textbf{Ingress} & \textbf{Egress} \\
\hline
Baseline & 4003.697 & 4003.697 & Baseline & 3113.987 & 3113.987 \\
D1-NC    & 4003.697 & 2342.916 & D1-NC    & 3113.987 & 1207.469 \\
D2-C     & 4003.697 & 1334.572 & D2-C     & 3113.987 & 1487.093 \\
U1-C     & 7612.005 & 4003.697 & U1-C     & 4639.221 & 3113.987 \\
U2-AGR   & 8037.091 & 4003.697 & U2-AGR   & 3609.695 & 3113.987 \\
U3-AGR   & 444.857  & 4003.697 & U3-AGR   & 444.857  & 3113.987 \\
\textbf{\sys}     & \textbf{444.857} & \textbf{1314.801} &
\textbf{\sys}     & \textbf{444.857} & \textbf{1188.774} \\
\hline
\end{tabular}
\label{table:traffic}
\vspace{-0.15in}
\end{table}
Figure~\ref{fig:com-time-aws} shows that applying existing network coding in the download phase (D1-NC) does not significantly reduce the average download time, but it does reduce the average waiting time by about 33\%. According to Figure~\ref{fig:split-time}, D1-NC makes the end of each client download phase closer. As the analysis in Section~\ref{sec:forwarding} shows, only the forwarding with the encoded blocks from the server is always valid, and the forwarding with the encoded blocks from other clients sometimes wastes the bandwidth.
We randomly select a communication round and count the validity of the encoded blocks received by each client. Figure~\ref{fig:invalid-nc} shows that up to 20\% of the encoded blocks received by the client are invalid, which not only wastes traffic but also delays the transmission of other valid encoded blocks. The benefits brought by coding and forwarding are offset by this invalid data transmission.
Therefore, the average download time of D1-NC is quite close to the baseline time.

When using our coding strategy in the download phase (D2-C), the client's forwarding efficiently uses the idle bandwidth through the client-to-client communication. The average download time is reduced by 60\% and the average waiting time is reduced by 33\% compared to the baseline. According to Figure~\ref{fig:split-time}, the download speeds of all clients have benefited from coding and the effective forwarding, especially the slower clients such as C5. In addition, Table~\ref{table:traffic} shows that server egress traffic is saved by 41\% (D1-NC) and 67\% (D2-C), respectively. Due to the coding mechanism, the data sent by the server is not duplicated, and the clients have more chances to get the global model partitions from their neighbors. Since each time forwarding is effective for D2-C, there are significant egress traffic savings in D2-C compared to D1-NC.

When our coding strategy is applied in the upload phase (U1-C), the average upload time is reduced by about 40\% and the average waiting time is reduced by about 32\%. 
When applying coded aggregation (U2-AGR/U3-AGR), the server decodes an aggregated global model from received blocks rather than per-client models; thus, their upload-phase time should be compared against the sum of upload and waiting time in other schemes. 
U2-AGR and U3-AGR achieve similar upload-phase communication time to U1-C, which is about 62\%-69\% of the communication time in the upload phase of the baseline. However, the ingress traffic of the server with U1-C and U2-AGR is relatively high because the algorithms require transmission redundancy for communication efficiency. The wait mode Coded-AGR (U3-AGR) can significantly reduce the ingress traffic through the hierarchical aggregation on encoded data blocks. Table~\ref{table:traffic} shows that the cost of accelerated communication in the upload phase of U1-C and U2-AGR is nearly twice as much server ingress traffic as in the baseline. The ingress traffic of U3-AGR is only about 11\% of the baseline.
With the full application of our coding strategies, \sys reduces the average download time by 59\% and the total communication time by 62\% compared to the baseline. 

In the North America topology, which is quite different from the global topology, the system has less communication heterogeneity as shown in Figure~\ref{fig:fluct-all}. Therefore, the download time, upload time, and waiting time of the baseline are small, which are only about 9s, 9.8s, and 5.4s, respectively. In this case, the advantage of existing network coding in the download phase (D1-NC) over the baseline is even smaller. Figure~\ref{fig:com-time-all} shows that our coding strategy in the download phase (D2-C) still achieves a 46\% improvement in the download time.
In the upload phase, our coding strategy with wait mode Coded-AGR (U3-AGR) reduces the upload communication time by 27\%  and decreases server ingress traffic to just 14\% of the baseline.
With the full application of our coding strategies, \sys reduces the total communication time by about 40\%.

\subsubsection{Effectiveness of Adaptive Redundancy Algorithm}\label{sec:traffic-dynamic}
To evaluate adaptive redundancy algorithm, we compare \sys under static vs.\ adaptive redundancy in terms of communication time and inter-client traffic. We set $\lambda=1.2$ by default.

\begin{table}[t]
\centering
\caption{The average traffic (Unit: MBytes)}
\vspace{-0.05in}
\begin{tabular}{|c|c|c|c|c|}
\hline
Topology & Role & Protocol & Ingress & Egress \\ \hline
\multirow{4}{*}{Global}
&\multirow{2}{*}{Server} & Static & 444.857 & 1314.801 \\ 
&&Adaptive & 444.857 & 1448.258 $\uparrow$ \\ \cline{2-5}
&\multirow{2}{*}{Client}&Static & 1656.407 & 1693.754 \\ 
&&Adaptive & 1564.140 \textcolor{darkgreen}{$\downarrow$} & 1586.658 \textcolor{darkgreen}{$\downarrow$}\\ \hline
\multirow{4}{*}{\makecell{North\\America}}
&\multirow{2}{*}{Server}&Static & 346.000 & 1235.715 \\ 
&&Adaptive & 346.000 & 1163.819 \textcolor{darkgreen}{$\downarrow$}\\ \cline{2-5}
&\multirow{2}{*}{Client}&Static & 1254.074 & 1501.217 \\ 
&&Adaptive & 942.353 \textcolor{darkgreen}{$\downarrow$} & 1124.019 \textcolor{darkgreen}{$\downarrow$}\\ \hline 
\end{tabular}
\label{table:traffic-dynamic}
\vspace{-0.15in}
\end{table}

For the global topology, Figure~\ref{fig:com-time-aws} shows that the communication time of \sys with the adaptive redundancy is 11\% lower than the time of \sys with static redundancy. In addition, the adaptive redundancy algorithm reduces communication traffic between clients. Table~\ref{table:traffic-dynamic} presents the average ingress and egress traffic for all clients, as well as the individual ingress and egress traffic for servers, under both static and dynamic redundancy. The results indicate that the dynamic mechanism reduces inter-client communication traffic by 6\%. For the North America topology, Figure~\ref{fig:com-time-all} shows that the communication time is comparable between the two schemes due to lower network heterogeneity and fewer fluctuations. The adaptive redundancy algorithm reduces the redundancy level dynamically, achieving up to a 25\% decrease in inter-client communication traffic.

\subsubsection{Numerical Experiments}
\begin{figure}[t]
    \centering
    \subfigure[Bandwidth Mean (Unit: Mbps)]{\includegraphics[width=0.24\textwidth, trim=0 0 0 15]{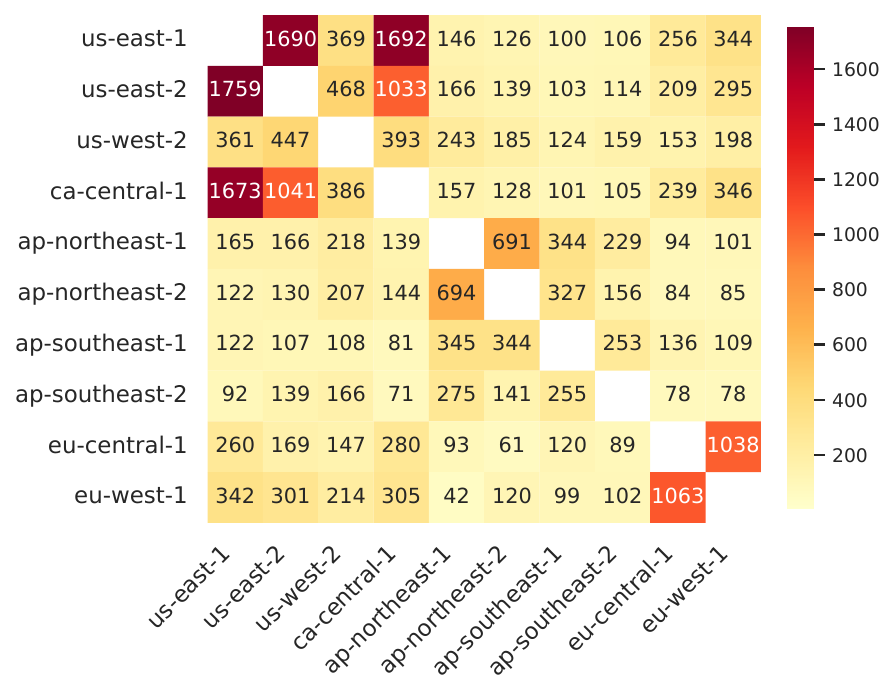}\label{fig:heat-mu}}
    \subfigure[Variance (Unit: Mbps\^{}2)]{\includegraphics[width=0.24\textwidth, trim=0 0 0 15]{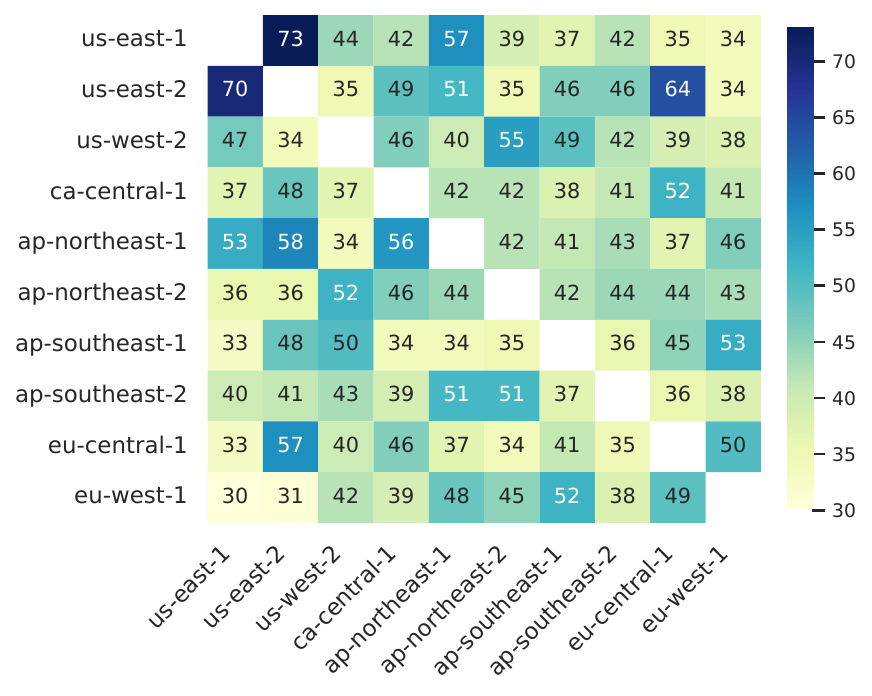}\label{fig:heat-sigma}}
    \vspace{-0.15in}
    \caption{The real-world distribution of the communication bandwidth between each pair of silos in the global topology.}
    \label{fig:heat}
    \vspace{-0.2in}
\end{figure}

We conduct numerical experiments to study the impact of the number of model partitions and how redundancy affects tolerance to network fluctuations.
We simulate \sys using the measured inter-silo bandwidth in the global topology (Figure~\ref{fig:heat}, mean and variance), and incorporate empirical distributions of training and coding time.

\begin{figure}[t]
    \centering
    \subfigure[Download $k$]{\includegraphics[width=0.2\textwidth, trim=0 10 0 10]{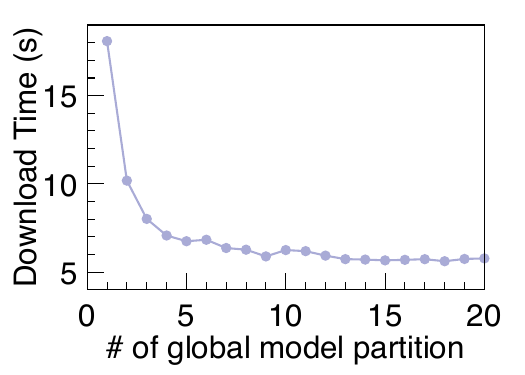}\label{fig:k-impact-download}}
    \subfigure[Upload $k$]{\includegraphics[width=0.2\textwidth, trim=0 10 0 10]{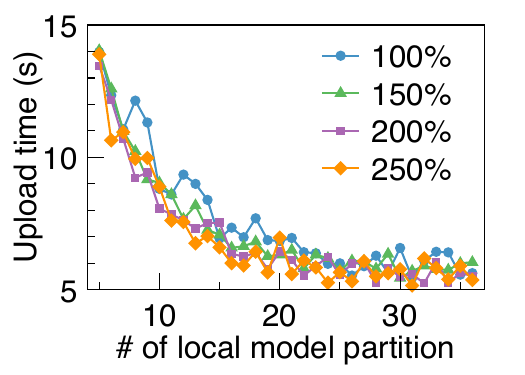}\label{fig:k-impact-upload}}
    \vspace{-0.1in}
    \caption{The impact of the number of model partitions $k$.}
    \vspace{-0.2in}
    \label{fig:k-impact}
\end{figure}

\begin{figure}[!t]
    \centering
    \subfigure[No faulty link]{\includegraphics[width=0.2\textwidth,trim=0 15 0 5]{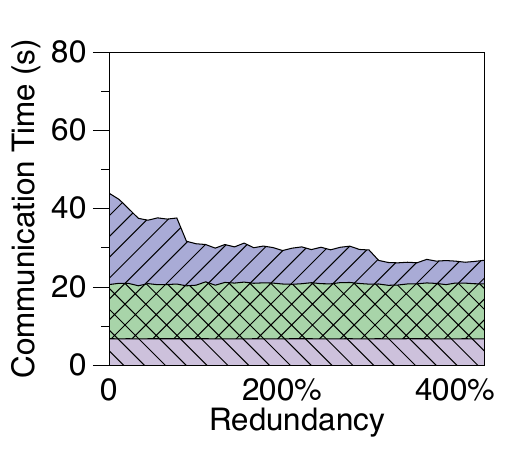}}\label{fig:fault-0}
    \subfigure[One faulty link]{\includegraphics[width=0.2\textwidth,trim=0 15 0 5]{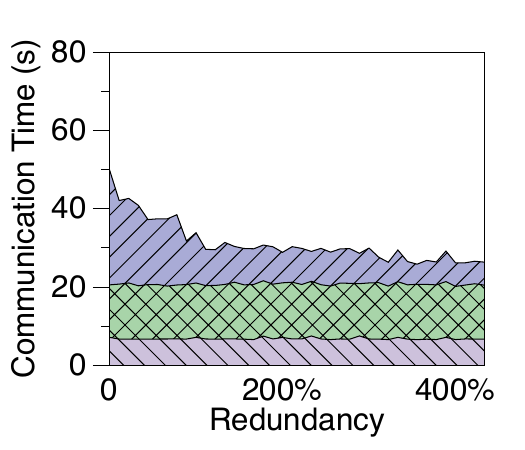}}\label{fig:fault-1}\\\vspace{-0.15in}
    \subfigure[Two faulty links]{\includegraphics[width=0.2\textwidth,trim=0 15 0 5]{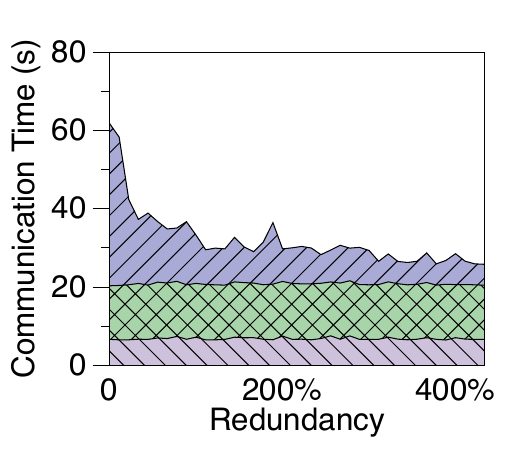}}\label{fig:fault-2}
    \subfigure[Three faulty links]{\includegraphics[width=0.2\textwidth,trim=0 15 0 5]{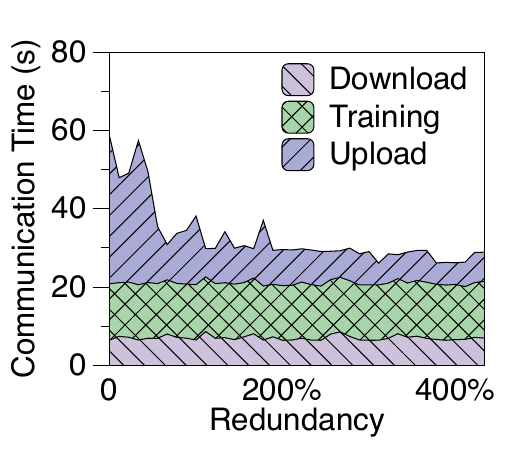}}\label{fig:fault-3}
    \vspace{-0.1in}
    \caption{The communication time for different redundancy and faulty links in the global topology.}
    \vspace{-0.2in}
    \label{fig:fault-aws}
\end{figure}
\noindent\textbf{Impact of the number of model partitions $k$}. In coding strategies, the number of model partitions $k$ is an important factor that affects the communication time. We study the effect of $k$ in the download and upload phases separately. 

For the download phase, Figure~\ref{fig:k-impact-download} shows that when $k=1$, the client can forward the data to its neighbors only after receiving the full global model and can hardly benefit from the coding, so the average download time is very close to the baseline. As $k$ grows, the model is partitioned at a finer granularity. The model partitions have more chances to be encoded and forwarded to the clients by the neighboring clients with good communication status, so the download time decreases. 
For the upload phase, we experiment with the impact of the number of local model partitions for four different redundancy scenarios (100\%, 150\%, 200\%, and 250\%). Figure~\ref{fig:k-impact-upload} shows that the upload time decreases as $k$ increases in four scenarios. The reason is that large $k$ gives the model partitions more opportunity to be aggregated and forwarded to the server by clients with fast transmission speeds. 
For both two phases, due to the inherent coding time of each model partition, the tendency to decrease stops when $k$ exceeds a threshold. Therefore, in practice, we should choose the appropriate $k$ based on the number of clients and the coding overhead.

\noindent\textbf{Impact of the coding redundancy.} 
Figure~\ref{fig:profile} shows that the real-world bandwidth fluctuations can lead to link failures (bandwidth at very low levels). We conduct experiments for the global topology under different numbers of faulty links to explore the relationship between redundancy and system tolerance. 
Figure~\ref{fig:fault-aws} shows that the performance improvement of increasing redundancy is significant even in the absence of link failures. 
This is because redundancy improves the utilization of idle bandwidth through client-to-client communication, and more data reaches the server through faster clients. As the number of faulty links increases, the system requires more and more redundancy to cope with the faulty links to maintain a stable communication efficiency. 

\subsubsection{Comparison with GDML communication}
We compare \sys with CREW~\cite{luo2024efficient} as a contextual reference, since CREW targets geo-distributed machine learning with a different synchronization model and is not a strict drop-in baseline for centralized cross-silo FL. 
Under the same global topology with \texttt{m5.8xlarge} instances, Table~\ref{tab:speedup_comparison} shows that CREW achieves a $2.13\times$ speedup over the baseline, while \sys achieves a $2.54\times$ speedup. 
This is because CREW relies on linear programming over profiled bandwidth status, whereas \sys uses coding to opportunistically adapt to bandwidth heterogeneity and fluctuations. 

\begin{table}[t]
\caption{Communication time and speedup of different protocols}
\label{tab:speedup_comparison}
\vspace{-0.05in}
\centering
\begin{tabular}{|p{1.5cm}|c|c|}
\hline
\textbf{Protocol} & \textbf{Communication time (s)} & \textbf{Speedup} \\
\hline
Baseline & 83.18 & 1$\times$ \\ \hline
CREW~\cite{luo2024efficient} & 39.04 & 2.13$\times$ \\ \hline
\textbf{FedCod} & \textbf{32.68} & \textbf{2.54}$\times$ \\ 
\hline
\end{tabular}
\vspace{-0.05in}
\end{table}
\begin{comment}
\subsubsection{Comparison to SOTA geo-distributed machine learning algorithm}
Recently, geo-distributed machine learning (GDML)~\cite{douillard2023diloco,luo2024efficient} has received increasing attention from researchers. Although the process of local training and parameter synchronization of GDML is similar to FL, most of the communication-efficient solutions for GDML are still coupled with training and aggregation algorithms, which cannot be applied in centralized cross-silo FL. CREW~\cite{luo2024efficient}, which works as an improved Partial Reduce / AllReduce communication protocol, although targeting a slightly different problem set, is still relevant and provides a useful baseline for comparison.

We conduct a comparison experiment in the global topology as mentioned in Section~\ref{sec:exp-setup} with \texttt{m5.8xlarge} instances. Since CREW~\cite{luo2024efficient} does not synchronize the parameters with the server, we can only count the total communication time per round for comparison. Table~\ref{tab:speedup_comparison} shows that compared to the baseline, CREW's speedup is $2.13\times$, while our \sys is up to $2.54\times$. The main reason is that CREW performs linear programming based on the profiled bandwidth status, while \sys uses coding to adapt to the network bandwidth in a more flexible way. Therefore, even though CREW does not synchronize the parameters with the server, its communication efficiency is still not as high as \sys.

\end{comment}

\begin{table}[t]
\centering
\caption{The testing accuracy of different communication protocols in different training stages}
\vspace{-0.05in}
\begin{tabular}{cccc}
\hline
\textbf{Protocol} & \textbf{Epoch=20} & \textbf{Epoch=40} & \textbf{Converge} \\ \hline
Baseline & 0.70 & 0.78 & 0.80 \\ 
HierFL & 0.70 & 0.79 & 0.80 \\ 
D1-NC & 0.73 & 0.79 & 0.80 \\ 
D2-C & 0.72 & 0.79 & 0.80 \\ 
U1-C & 0.70 & 0.79 & 0.80 \\ 
U2-AGR & 0.70 & 0.79 & 0.80 \\ 
U3-AGR & 0.70 & 0.79 & 0.80 \\ 
\sys & 0.70 & 0.79 & 0.80 \\ 
Adaptive & 0.70 & 0.78 & 0.80 \\ 
\hline
\end{tabular}
\label{table:conformance}
\vspace{-0.15in}
\end{table}

\subsubsection{Conformance Experiments}
Our proposed algorithms for cross-silo FL, including the coding strategies and the adaptive redundancy algorithm, are lossless communication protocols. Therefore, they do not affect the learning convergence of the global model. Table~\ref{table:conformance} shows that our coding strategies in both the download and upload phases do not affect the convergence of the global model. The global models trained in the systems with coded aggregation and adaptive redundancy algorithm can also converge to the test accuracy with the same convergence rate and the same accuracy level.

\section{Related Work}
\subsection{Communication-Efficient FL}

Communication-efficient FL methods fall into two categories: \textit{algorithmic solutions} that trade convergence or accuracy for communication efficiency, and \textit{network optimizations} decoupled from FL algorithms.

\textit{Algorithmic solutions.} Most existing communication-efficient FL methods attempt to address the communication problem from an algorithmic perspective. To improve the communication efficiency, \textit{they rely on specific FL setups and cannot be applied to generic FL algorithms}. 
Some previous work increases the number of local training iterations between communication rounds and decreases the frequency of communication~\cite{mcmahan2017communication,stich2018local,luo2021cost}. 
While lower communication frequencies reduce the amount of data exchanged, they are more likely to introduce model drift~\cite{karimireddy2020scaffold}.
Quantization and compression techniques in FL also reduce the amount of communication per round~\cite{rothchild2020fetchsgd,dorfman2023docofl}, but the loss of information due to lossy compression can affect convergence~\cite{liu2022hierarchical,sun2024convergence}. 
Considering that communication delays are usually caused by unstable connections between server and clients, some studies propose one-shot~\cite{zhang2022dense} or decentralized FL~\cite{tang2018d,marfoq2020throughput}, which modify the centralized architecture and aggregation algorithm. However, modifications in the aggregation algorithms always affect the learning convergence~\cite{sun2023mode}.

\textit{Network optimizations.} From a network perspective, only a few studies use hierarchical aggregation to address the system heterogeneity and save communication resources~\cite{abad2020hierarchical,wang2021resource,xu2022hierfedml}.
They divide the clients into clusters based on the network topology and geographic proximity, and perform early aggregation in the cluster. The early aggregated models are uploaded from the clusters to the server for final aggregation. Clients in the same cluster often have more stable network connections to each other, and early aggregation in the group can reduce bandwidth consumption, speeding up the upload process. 
This paradigm is more appropriate for cross-device scenarios where most clients are in the same Local Area Network (LAN) or base station coverage area. While in the cross-silo scenario, there are fewer opportunities to leverage this type of ``locality'' due to the geo-distributed nature. 
Instead, if the transmission speed between the clients performing early aggregation is not fast enough, the efficiency will be worse than traditional direct server-to-client communication. In contrast, \sys uses coding strategies to better adapt to cross-silo FL networks in WAN environments, achieving more efficient communication with the same model quality.

\subsection{Network Coding}
Network coding improves multicast throughput and robustness by coding packets at intermediate nodes~\cite{ahlswede2000network}, and has been widely used in peer-to-peer content distribution~\cite{gkantsidis2005network}. Random linear network coding uses random coefficients to mitigate packet loss and volatility in such networks~\cite{ho2006random,deb2006algebraic,li2011random}. Some recent attempts apply coding ideas to FL but consider limited system settings (\eg, few clients or specific topologies) and do not generalize to practical multi-client cross-silo deployments~\cite{kong2020network}. Our work builds a general protocol that integrates coding into cross-silo FL dataflow.

\subsection{Coded Computing}
Unlike communication redundancy (\eg, network coding in FL), coded computing introduces \emph{computational} redundancy to mitigate stragglers~\cite{ye2018communication,yun2023slimfl}. These methods typically code matrix multiplications across clients, requiring data overlap or scaling poorly to large neural networks. In contrast, we add \emph{communication} redundancy to exploit idle bandwidth in the system, remaining decoupled from the FL algorithms and transparent to the learning process.

\section{Conclusion}
This paper proposes \sys, a new application layer communication protocol designed to improve communication efficiency in cross-silo FL. By using coding strategies specific to the different characteristics of the data flow of the upload and download phases, \sys makes efficient use of the available bandwidth and accelerates the transfer of model weights through client-to-client communication.
Furthermore, the adaptive redundancy algorithm in \sys dynamically balances redundancy levels in response to network fluctuations, ensuring system resilience while minimizing unnecessary communication traffic.
Extensive experiments demonstrate that \sys significantly reduces communication time without compromising learning convergence.

\section*{acknowledgments}
This work was supported by the National Key R\&D Program of China (2022YFB4402102), National Natural Science Foundation of China (NO. 62472284), and Shanghai Key Laboratory of Scalable Computing and Systems. The work of Hao Wang was supported in part by NSF 2534286, 2523997, and 2315612.

\bibliographystyle{IEEEtran}
\bibliography{main.bib}

\end{document}